\documentstyle[12pt,epsf]{article}

\begin{document}
\def\beq{\begin{equation}}
\def\eeq{\end{equation}}
\def\beqa{\begin{eqnarray}}
\def\eeqa{\end{eqnarray}}
\def\noin{\noindent}
\def\grad{{\bf \nabla}}
\def\bv{{\bf v}}
\def\bB{{\bf B}}
\def\bJ{{\bf J}}
\def\bE{{\bf E}}
\def\bK{{\bf K}}
\def\bx{{\bf x}}
\def\pa{\partial}
\def\eps{\epsilon_{\alpha\beta}}
\def\rta{\rightarrow}
\def\lra{\leftrightarrow}
\def\tm{\tilde{m}}
\def\tn{\tilde{n}}
\def\tbv{\tilde{\bv}}
\def\tbK{\tilde{\bK}}
\def\tP{\tilde{P}}
\def\trho{\tilde{\rho}}
\def\hm{\hat{m}}
\def\hn{\hat{n}}
\def\hbv{\hat{\bv}}
\def\hbK{\hat{\bK}}
\def\hP{\hat{P}}
\def\hrho{\hat{\rho}}
\def\cE{{\bf {\cal E}}}
\def\cB{{\bf {\cal B}}}
\def\bbJ{\bar{\bJ}}
\def\brho{\bar{\rho}}
\def\bsig{\bar{\sigma}}
\def\ccB{{\cal B}}
\def\ccE{{\cal E}}
\def\ccA{{\cal A}}
\def\hcA{\hat{\ccA}}
\titlepage
\begin{flushright} QMW-PH-96-02
\end{flushright}
\vspace{3ex}
\begin{center} \bf
{\bf DUALITY INVARIANT MAGNETOHYDRODYNAMICS AND DYONS}\\
\rm

\vspace{3ex}

Omduth Coceal$^{a}$\footnote{e-mail: o.coceal@qmw.ac.uk},
Wafic A. Sabra$^{b}$\footnote{e-mail:
uhap012@vax.rhbnc.ac.uk} and
Steven Thomas$^{a}$\footnote{e-mail: s.thomas@qmw.ac.uk}\\
\vspace{2ex}
$^{a}${\it Department of Physics\\
Queen Mary and Westfield College\\
Mile End Road\\
London E1 4NS\\
U.K.}\\
\vspace{4ex}
$^{b}${\it Department of Physics\\
Royal Holloway and Bedford New College\\
Egham\\
Surrey TW20 0EX\\
U.K.}\\
\vspace{4ex}
ABSTRACT
\end{center}
\noindent
The theory of magnetohydrodynamics is extended
to the cases of a plasma of separate magnetic and
electric charges, as well as to a plasma of  dyons
respectively. In both these cases the
system possesses electric-magnetic duality symmetry.
In the former case we find that because of
the existence of
two independent generalized Ohm's law equations, the limit
of infinite
electric and magnetic conductivity results in the vanishing
of both
electric and magnetic fields, as well as the corresponding
currents. In the dyonic  case,  we find  that the
resulting duality-invariant system of equations are
equivalent to those of ordinary MHD, after suitable field  
redefinitions.
\newpage
Recently Polyakov [1] presented a new radical approach to
the study of two dimensional (2d) turbulent fluids by proposing that  
certain
conformal
field theories (CFT)  might offer a description of fully developed  
turbulence.
Ferretti and Yang  [2] applied these ideas to the effective field  
theory
approximating
2d plasmas- $i.e.,$ 2d magnetohydrodynamics (MHD).
Recently we presented some new results [3]  which  extended the
work of [2].  Strictly speaking, the theories studied in [2,3]
were 3d MHD  dimensionally reduced to 2d-$i. e.$ all fields are taken  
to be
independent
of the third coordinate. In this way a generalized 2d MHD is obtained  
in which
there are passive scalars as well as the usual velocity, electric and  
magnetic
fields.
Pure 2d MHD can be obtained by consistently setting to zero the  
passive
scalars.

In this letter, motivated by a desire to obtain even more generalized  
MHD
equations in 2d, with a view to applying CFT techniques in  looking  
for
possible solutions in the turbulent regime [4], we shall construct  
the analogue
of the MHD
equations  of plasmas that incorporate magnetic charges.
Of course such theories like ordinary  MHD, should find much broader
physical applications.\footnote{For example,
there has been interesting recent work concerning the existence of  
electric
string solutions in the MHD of magnetic monopole plasmas  [5] } As we  
shall see
there are at least three ways of accomplishing this. One could  
consider
a purely magnetic plasma, consisting of  oppositely charged  magnetic
monopoles.
This is in fact the direct analogue of the usual plasma of electric  
charges
(see
 e.g. [5]).
Secondly we can consider a plasma where both types of charges are  
present-
a kind of interacting pair of electric and magnetic plasmas.
Finally, we may consider plasmas involving particles that carry both
electric and magnetic charges- i.e. dyons. We shall see that all  
these
possibilities
can display MHD-like equations  having different physical properties.
Moreover all these possibilities are deeply connected with the idea  
of
electric-magnetic duality.

First we consider the situation where we have a plasma
consisting of
magnetic
monopoles and their ionic counterparts - ``magneto-ions".
Following the well known methods (see for example [6])
 that yield ordinary MHD in
3 dimensions, by deriving 1-fluid equations
from the 2-fluid  description, we shall obtain analogous
``dual" MHD equations (referred to perhaps more appropriately as
electrohydrodynamics in [5]).  We begin with
Maxwell's  equations  including both electric and magnetic currents
$\bJ_{e}  , \bJ_{m}$
\newpage
\beqa\label{eq:3.1}
\grad\cdot\bE & = & \rho_{c},  \qquad\qquad\qquad\quad \grad\cdot\bB   
=
\rho_{m}, \cr
\grad\times\bB & = & \frac{4\pi}{c}\bJ_{e} + \frac{1}{c}\pa_{t}\bE,  
\qquad
\grad\times\bE  =  -\frac{4\pi}{c}\bJ_{m} - \frac{1}{c}
\pa_{t}\bB
\eeqa
which are well known to be
 invariant under the electric-magnetic duality transformations (see  
for example
[7]):
\beqa\label{eq:2.1}
\bE\rta\bB,\quad \bB\rta -\bE  \quad
\bJ_{e}\rta\bJ_{m},\quad  \bJ_{m}\rta -\bJ_{e} \quad
\rho_{c}\rta\rho_{m},\quad  \rho_{m}\rta -\rho_{c}
\eeqa
where $\rho_{m}$ and $\rho_{e}$ are the
corresponding magnetic  and electric charge densities
respectively.

We now show how we can implement and extend this symmetry
to the full set of MHD equations.
The  two-fluid equations for our monopole plasma analogous to
ordinary electric plasmas
are:
\beqa\label{eq:2.2}
\pa_{t}\tilde{n}_{s}(\bx,t)+\grad\cdot(\tilde{n}_{s}\tilde{\bv}_{s})  
& = & 0,
\cr
\tilde{m}_{s}\tilde{n}_{s}(\pa_{t}\tilde{\bv}_{s}+\tilde{\bv}
_{s}\cdot\grad\tilde{\bv}_{s}) & = & -\grad \tilde{P}_{s}
+g_{s}\tilde{n}_{s}(\bB-\frac{1}{c}\tilde{\bv}_{s}\times\bE)+
\tilde{\bK}_{s}(\bx),
\eeqa
where the quantities $\tm_s ,  g_s , \tP_s $ and $\tn_s $ are  
respectively the
mass,
magnetic charge, pressure and number density of the individual  
species $s$,
where $s = g $ or $i$ corresponding to the monopoles or magnetic  
ions,
$\tbv_s $ is the corresponding velocity field and $\tbK_s $   
represents the
change in momentum of
species $s$ at position $\bx $ due to collisions. Usual  scattering
arguments [6] lead us to take $\tbK_i = - \tbK_g $. Note that
$g_g = - g_i = g $ is the
magnetic
charge of the monopole.   The force
equation is  the dual of the standard one, using
(\ref{eq:2.1}).

\noin
In order to sum up over species,  the
following quantities are defined,
\beqa\label{eq:4}
\tilde{\rho}_{M}(\bx) & \equiv &
\tilde{m}_{g}\tilde{n}_{g}(\bx)+\tilde{m}_{i}\tilde{n}_{i}(\bx),  
\qquad
\rho_{m}(\bx)  \equiv  -g(\tilde{n}_{g}-\tilde{n}_{i}), \cr
\tilde{\bv} & \equiv &
\frac{1}{\tilde{\rho}_{M}}(\tilde{m}_{g}\tilde{n}_{g}\tbv_{g} +
\tilde{m}_{i}\tilde{n}_{i}\tilde{\bv}_{i}), \qquad
\bJ_{m}  \equiv  g(\tilde{n}_{i}\tilde{\bv}_{i} -
\tilde{n}_{g}\tilde{\bv}_{g}), \cr
\tilde{P} & \equiv & \tilde{P}_{g} + \tilde{P}_{i}.
\eeqa
If we  assume that $\tilde{m}_{g}\ll\tilde{m}_{i}$,
$\tilde{n}_{g}\approx\tilde{n}_{i},$
 we are then led to
the following
one-fluid equations,
\beqa\label{eq:2.4}
\pa_{t}\tilde{\rho}_{M}+\grad\cdot
(\tilde{\rho}_{M}\tilde{\bv}) & = & 0, \qquad
\pa_{t}\rho_{m}+\grad\cdot \bJ_{m}  =  0, \cr
\tilde{\rho}_{M}(\pa_{t}\tilde{\bv} +
\tilde{\bv}\cdot\grad\tilde{\bv}) & = &
-\grad \tilde{P} + \rho_{m}\bB
- \frac{1}{c}\bJ_{m}\times\bE, \cr
\frac{\tilde{m}_{g}\tilde{m}_{i}}{\rho_{M}g^{2}}\pa_{t}\bJ_{m
} & = &
\frac{\tilde{m}_{i}}{2\rho_{M}g}\grad \tilde{P}
+ (\bB-\frac{1}{c}\tilde{\bv}\times\bE) +
\frac{\tilde{m}_{i}}{\rho_{M}gc}\bJ_{m}\times\bE
- \frac{\bJ_{m}}{\sigma_{m}}
\eeqa

\noin
coupled to the duality invariant Maxwell's equations of  
(\ref{eq:3.1})
with $\rho_c = \bJ_e = 0 $. The last equation in (5) is the magnetic  
analogue
of
Ohm's law, with magnetic conductivity $\sigma_m $.

Finally we note that these ``electrohydrodynamic" equations [5] are  
related to
the usual
MHD equations through the
duality transformations (\ref{eq:2.1}), together with  the
transformations $
e\rta g, \,  g\rta -e ,\, m_{e}\lra\tilde{m}_{g} , \,m_{i}\lra  
\tilde{m}_{i},\,
\sigma_{e}\lra\sigma_{m} ,\, \rho_{M}\lra\tilde{\rho}_{M} $ and $
\bv  \lra  \tilde{\bv} $.

We now turn to the magnetohydrodynamic description of a
mixed plasma
consisting of electrons, ions, magnetic monopoles and their
corresponding
magnetic ``ions". For such a system the relevant equations
have to be
amended to
take into account the presence of the different species.
Maxwell's
equations,
for example, have to be generalized to include the effects
of both electric
and magnetic charges and currents in a duality-invariant
way, hence  the self-dual
equations (\ref{eq:3.1}) are relevant.

In this case we also have two sets of two-fluid
conservation and force
equations:
\beqa\label{eq:3.2}
\pa_{t}n_{s}+\grad\cdot(n_{s}\bv_{s}) & = & 0 ,~~~s=e,i,\cr
\pa_{t}\tilde{n}_{s}+\grad\cdot(\tilde{n}_{s}\tilde{\bv}_{s})
 & = & 0 ,~~~s=g,i, \cr
m_{s}n_{s}(\pa_{t}\bv_{s}+\bv_{s}\cdot\grad\bv_{s}) & = &
-\grad P_{s}
+q_{s}n_{s}(\bE+\frac{1}{c}\bv_{s}\times\bB)+\bK_{s}, \cr
\tilde{m}_{s}\tilde{n}_{s}(\pa_{t}\tilde{\bv}_{s}+\tilde{\bv}
_{s}\cdot\grad\tilde{\bv}_{s}) & = & -\grad \tilde{P}_{s}
+g_{s}\tilde{n}_{s}(\bB-\frac{1}{c}\tilde{\bv}_{s}\times\bE)+
\tilde{\bK}_{s},
\eeqa
where $q_{e}=-e, q_{i}=e$ and $g_{g}=-g, g_{i}=g$, $\bv_s  , n_s $  
and
$P_s $ are the electric analogues of the quantities defined earlier.

After summing over species, the following
one-fluid conservation
equations are obtained:
\beqa\label{eq:3.4}
\pa_{t}\hat{\rho}_{M}+\grad\cdot(\hat{\rho}_{M}\hat{\bv})
=  0, \quad
\pa_{t}\rho_{c} + \grad\cdot\bJ_{e} =  0, \quad
\pa_{t}\rho_{m} + \grad\cdot\bJ_{m} =  0
\eeqa
where the quantities $\hat{\rho}_M , \hat{\bv} , \rho_c $ and $\rho_m  
$ are
again
just the generalizations of similar quantities in (\ref{eq:4})
to the 4-species case.
The  generalized Navier-Stokes equation is:
\beq\label{eq:3.5}
\hat{\rho}_{M}[\pa_{t}\hat{\bv} +
(\hat{\bv}\cdot\grad)\hat{\bv}]=-\grad
\hat{P} + \rho_{c}\bE
+ \rho_{m}\bB + \frac{1}{c}\bJ_{e}\times\bB -
\frac{1}{c}\bJ_{m}\times\bE
+ \sum_{s}(\bK_{s}+\tilde{\bK}_{s}).
\eeq
$\hat{P} $ being the total pressure and $\bK_s $ the collision
term that changes the momentum of electric charges. Evidently
(8) is invariant under electric-magnetic duality.

Previously we considered a plasma
of purely magnetic charges, (in analogy with pure electric)
where standard arguments  led to the
vanishing of $\sum_{s} {\tbK}_s $ (or
$\sum_{s } {\bK}_s $) respectively.  However, in the present  
situation
where both types
of charges are present,
we have to be a little more cautious as there could, in
principle,  be
interactions between electric
and magnetic charges that can  influence the sum $\sum_{s}
{\bK}_s +
{\tilde {\bK}}_s $. To understand
this problem further, it is useful to consider 2-dimensions,  and  
recall some
well
known
properties  [8]  of  Coulomb gases consisting of
both electric and magnetic charges.
  What is evident from [8]
is that   magnetic and electric charges  couple only
through  Bohm-Ahranov [9] type interactions of the form
$\sum_{i \neq j}^N e_i \Phi (z_i - y_j ) g_j $. Here
we consider $N$ electric and magnetic charges
$e_i , g_j ,$ located at the points $z_i , y_j $ of the complex  
plane,
and $\Phi (z) = Arg(z) $. This topological kind of interaction
is also present in 3 dimensions [10].
 It is clear  that
whether in 2 or 3 dimensions,
 $\bK_s $( $\tilde{\bK}_s $) are predominantly influenced by the
effective Coulomb
forces
which act between charges  of the same type be they electric  or  
magnetic.
Irrespective of
the particular
nature of such interactions,
what is important is that in our system the effective
forces of collision will occur  between charges of
the same type. This  leads to a picture of two
plasmas, one with
electric the other with magnetic charge carriers whose interaction
 is weak compared to their self interaction.
 Following  arguments similar to those discussed earlier,
we then expect that
collisions between electrons and their corresponding ions
will give $\sum_{s}
\bK_s \, = \, 0 $
and similarly for the magnetic charge carriers, $\sum_{s}
\tilde{\bK}_s \, = \, 0$ . Consequently,
these terms drop out of the 1-fluid force equation (\ref{eq:3.5})  
above.
\bigskip

\noin
In principle we can now construct two separate Ohm's law
equations
corresponding
to $\bJ_{e}$ and $\bJ_{m}$ respectively:
\beqa\label{eq:3.7}
\pa_{t}\bJ_{e} & = & -\frac{e}{m_{i}}\grad P_{i} +
\frac{e}{m_{e}}\grad P_{e}+e^{2}(\frac  
{n_{e}}{m_{e}}+\frac{n_{i}}{m_{i}})\bE +
\frac{e^{2}n_{e}}
{m_{e}c}\bv_{e}\times\bB +
\frac{e^{2}n_{i}}{m_{i}c}\bv_{i}\times\bB
\cr
&   & \mbox{} + \sum_{s}\frac{q_{s}}{m_{s}}\bK_{s}, \cr
\pa_{t}\bJ_{m} & = & -\frac{g}{\tm_{i}}\grad \tP_{i} +
\frac{g}{\tm_{g}}\grad
\tP_{g}
+g^{2}
(\frac{\tn_{g}}{\tm_{g}}+\frac{\tn_{i}}{\tm_{i}})\bB -
\frac{g^{2}\tn_{g}}
{\tm_{g}c}\tbv_{g}\times\bE -
\frac{g^{2}\tn_{i}}{\tm_{i}c}\tbv_{i}\times\bE
\cr
&   & \mbox{} + \sum_{s}\frac{g_{s}}{\tm_{s}}\tbK_{s}.
\eeqa
However to proceed further, we need to make a number of
approximations and
simplifying assumptions to obtain Ohm's laws that are written in  
terms
of 1-fluid quantities only. The only new assumptions we need make
in addition to those already used up to this point (and suitably  
generalized
to the present case), are that $| \bv_e | \gg |\bv_i | $ and
$|\tbv_g | \gg | \tbv_i | $, which is natural
for a plasma of such charges in thermal equilibrium when we recall  
that
$m_e \ll m_i $ and $\tm_g \ll \tm_i $. After some calculation we  
arrive at the
1-fluid formulation for the generalized Ohm's laws
\beqa\label{eq:3.19}
\frac{m_{e}m_{i}}{\hrho_{M}e^{2}}\pa_{t}\bJ_{e} & = &
\frac{m_{i}}{4\hrho_{M}e}\grad\hP
+ \frac{1}{2}\bE+\frac{1}{c}\hbv\times\bB -
\frac{m_{i}}{\hrho_{M}ec}\bJ_{e}\times\bB
- \frac{1}{2\sigma_{e}}\bJ_{e} \cr
\frac{\tm_{g}\tm_{i}}{\hrho_{M}g^{2}}\pa_{t}\bJ_{m} & = &
\frac{\tm_{i}}{4\hrho_{M}g}\grad\hP
+ \frac{1}{2}\bB-\frac{1}{c}\hbv\times\bE +
\frac{\tm_{i}}{\hrho_{M}gc}\bJ_{m}\times\bE
- \frac{1}{2\sigma_{m}}\bJ_{m}
\eeqa
(\ref{eq:3.1}), (\ref{eq:3.5}) and (\ref{eq:3.19}) constitute the  
equations of
duality invariant MHD of a plasma of $e$ and $g$ charges.
Before moving on to consider an alternative magnetic -electric plasma
description
where the two types of charges are localized at a point, thus  
defining a dyon,
we shall
investigate the ideal limit of
the duality invariant
MHD equations
(\ref{eq:3.1}), (\ref{eq:3.5}) and (\ref{eq:3.19}).  Studying  
infinitely
conducting
 (i.e. $\sigma_e \rightarrow \infty $)
ordinary (electric) plasmas,  apart from being a useful
simplifying limit, has proved useful in exhibiting many
interesting
properties of these
systems [6]. Furthermore, when one considers the
2-dimensional case, the
direct relevance of
conformal field theory in describing certain properties of
the turbulent
phase is clearer in this
limit [2,3], although some recent work has also considered
finite conductivity [3] in this context.

To obtain the equations of ideal MHD,  we shall adopt the
usual
approximations [6] of low
temperature, so the pressure term $\grad \hat {P} $ can be
neglected, and
low frequency, so all fields with time derivatives can be dropped
compared to those
without. In particular if
we apply these approximations to the generalized Ohm's law
equations (\ref{eq:3.19}),
and further assuming
that one may neglect the $\bJ_{e}\times\bB$ term compared
to
$\hbv\times\bB$ [6], as well as
the magnetic analogue of this (which is neglecting $\bJ_{m}\times  
\bE$
with respect to $\hbv\times\bE$), we obtain
\beqa\label{eq:3.20}
\bJ_{e} =\sigma_e (\bE + \frac{2}{c} \hbv \times
\bB), \qquad \qquad \bJ_{m} = \sigma_m (\bB -
\frac{2}{c} \hbv \times \bE).
\eeqa

We see that the above approximations still imply that both
equations in (\ref{eq:3.20}) are
related under the duality transformations of eqs (\ref{eq:2.1}).  Now
in the limits
$\sigma_e \rightarrow\infty$ and $\sigma_m\rightarrow\infty$ the  
condition
for there to exist
both finite
magnetic and electric currents requires
\beqa\label{eq:3.21}
\bE = -\frac{2}{c} \hbv \times \bB, \qquad
\bB =\frac{2}{c} \hbv \times \bE.
\eeqa
Again equivalence of these equations under duality is
obvious.
Equations (\ref{eq:3.1}), (\ref{eq:3.5}) and (\ref{eq:3.21})  
constitute the
 ideal, duality
invariant
equations  MHD for a plasma of separate electric and
magnetic charges.
Note that for ordinary electric plasmas in this limit,
since there is a single
Ohm's law [6] only the first equation in (\ref{eq:3.21}) is
applicable. Having said that,
the physical consequences of having two generalized Ohm's
law eqns. (\ref{eq:3.19})
instead of one are dramatic. This is because it is easy to
show that (\ref{eq:3.21})
constitute an overdetermined set of equations for $\bE $ and
$\bB $,
so that their only solution is $\bE \, = \, \bB \, = \, 0
$. In the case of a single
Ohm's law, e.g. the case of a purely electric plasma, the
first of eqns (\ref{eq:3.21})
would simply determine $\bE$ in terms of the velocity and
magnetic fields.
 Ultimately we can understand the origin of
two Ohm's laws in our mixed plasma as a consequence of
both electric and magnetic charges coupling to the {\it same }
$\bE $ and $\bB $ fields. Since the latter are obviously coupled
via Maxwell's eqns, these two plasmas are never really decoupled.

 %Inserted by TeXtelmExtel

The duality invariant Maxwell equations with vanishing $\bE
$ and
$\bB $ fields require that the corresponding currents
$\bJ_m $
and $\bJ_e $ vanish. One is tempted to interpret this  as
confinement
of both electric and magnetic charge in the plasma, which
is reflected in the
fact that the duality invariant MHD equations simply reduce to those  
of
an incompressible,
inviscid fluid when one has vanishing $\bE $ and $\bB$
fields.
However, such a  conclusion should be treated with caution.
In deriving the
1-fluid equations describing generalized Ohm's laws for
both $\bJ_e $
and $\bJ_ m $, we used the  approximation that $ {\bf v}_i
\ll {\bf v}_e $
and $ \tilde{{\bf v}}_i \ll \tilde{{\bf v}}_e $, which as
we noted was quite
natural given the hierarchical values of the masses $m_e ,
m_i ,
{\tilde{m}}_g, $ and ${\tilde{m}}_i $ as long as the system
is
completely in  the ${\it plasma} $ state $i.e.$, we have a
fluid of  separate
charges  {$\pm e, \pm g $}. Thus as the system approaches
confinement
of both kinds of charge, the 1-fluid formalism breaks down.
This is perhaps
not surprising since information about the plasma has been
lost
not only in moving from the 4-fluid (recall we now have
4 species of charge carriers) to the 1-fluid description,
but equally in choosing the low frequency, long wavelength
approximation.
Certainly one might expect short distance effects to be
important as
electron-ion and magnetic monopole-magneto ions recombine.

 %Inserted by TeXtelmExtel

In the final part of this letter we shall
consider a dyonic plasma, $i. e.$ one in which there
are {\em two}
species of dyons with opposite electric and magnetic
charges. Recall that dyons (see for example the review [7])
 are particles that carry
both electric and magnetic charge. The first dyon
species is the analogue of electrons and the second type
(``dyonic ions")
is analogous to ions. Let $m_{d}, m_{di}$ be the
corresponding masses with
$m_{d}\ll m_{di}$.
Their respective charges are defined as follows:
$q_{s}= -e $ if $ s=d$ and $q_s = +e $ if $ s=di $ .
Similarly, $g_{s}= -g $ if $ s=d$ and $g_s = +g $ if $ s=di $.

 %Inserted by TeXtelmExtel

The plasma equations for species $s$ will be:
\beqa\label{eq:4.1}
\pa_{t} n_{s}(\bx , t) + \grad\cdot (n_{s} \bv_{s}) & = & 0
\cr
m_{s}n_{s} (\pa_{t}\bv_{s} + \bv_{s}\cdot\grad \bv_{s}) & =
& -\grad P_{s}
+ q_{s}n_{s}(\bE+\frac{1}{c}\bv_{s}\times\bB), \cr
&  & \mbox{} +
g_{s}n_{s}(\bB-\frac{1}{c}\bv_{s}\times\bE)+\bK_{s},
\eeqa
along with the dualized Maxwell equations
(\ref{eq:3.1}).

For the dyonic fluid, the electric and magnetic currents
are defined as
\beqa\label{eq:4.3}
\bJ_{e}\equiv
\sum_{s} q_{s}n_{s}\bv_{s}
=-e(n_{d}\bv_{d}-n_{di}\bv_{di}),
\quad\bJ_{m}\equiv  \sum_{s} g_{s}n_{s}\bv_{s}
=-g(n_{d}\bv_{d}-n_{di}\bv_{di}).
\eeqa
Hence, for the one-fluid dyonic plasma,  $\bJ_{e}$
and $\bJ_{m}$ are related by
\beq\label{eq:4.4}
\bJ_{e}=\frac{e}{g}\bJ_{m}.
\eeq
This relation is  crucial and has  important
implications.
Usual manipulations  yield the conservation equations:
\beqa\label{eq:4.5}
\pa_{t}\rho_{M}'+\grad\cdot (\rho_{M}'\bv)  =  0,\quad
\pa_{t}\rho_{c}+\grad\cdot \bJ_{e}  =  0,\quad
\pa_{t}\rho_{m}+\grad\cdot \bJ_{m} =  0,
\eeqa
with $\rho_{M'} $ and $\bv $ the usual 1-fluid quantities, and
similar to the connection between electric and magnetic currents,
one may show that $
\rho_{c}=\frac{\displaystyle e}{\displaystyle g}\rho_{m} $.
The one-fluid force equation for a dyonic plasma is
\beqa\label{eq:4.8}
\rho_{M}' (\pa_{t}\bv+\bv\cdot\grad \bv)  =  -\grad P' +
\rho_{c}\bE + \rho_{m}\bB\mbox{}+ \frac{1}{c} \bJ_{e}\times\bB -  
\frac{1}{c}
\bJ_{m}\times\bE,
\eeqa
$P' $ being the total pressure.
This defines the duality-invariant force equations for a
dyonic fluid.

Since the electric and magnetic currents are related for
dyons, this means that the dyonic
 generalized Ohm's law must be self dual. Making standard  
approximations
we are led to the following form of the Ohm's law for
the dyonic plasma
\beqa\label{eq:4.15}
\frac{m_{d}m_{di}}{\rho_{M}'e^{2}}\frac{\pa\bJ_{e}}{\pa t}
& = &
\frac{m_{d}}{2\rho_{M}'e}\grad P'
+ (\bE+\frac{1}{c}\bv\times\bB) +
\frac{g}{e}(\bB-\frac{1}{c}\bv\times\bE)
\cr
 &   & \mbox{} - \frac{m_{d}}{\rho_{M}'ec}\bJ_{e}\times\bB
+ \frac{m_{d}}{\rho_{M}'ec}\bJ_{m}\times\bE -
\frac{1}{\sigma_{e}}\bJ_{e},
\eeqa
where the electric conductivity $\sigma_{e} $ appears
through the Taylor expansion of  $\bK_{di} $ in terms of the
relative velocities $(\bv_d - \bv_{di})$
\beqa\label{eq:4.12}
\bK_{di}  \approx C\bJ_{e} =  
-\frac{\rho_{M}'e}{m_{di}\sigma_{e}}\bJ_{e}.
\eeqa
 One could equally express $\bK_{di} $ in terms of $\bJ_m $
and a magnetic conductivity $\sigma_m $, which by virtue of
(\ref{eq:4.4}) is given as
$\sigma_m = \frac{\displaystyle g^2}{\displaystyle e^2} \sigma_e.$

This completes our derivation of the dyonic one-fluid
equations. They can be
expressed in a manifestly duality-invariant way by a
suitable redefinition
of the fields.
Define duality-invariant electric and magnetic fields
\beqa\label{eq:4.18}
\cE  =  e\bE+g\bB, \qquad\cB  =  g\bE-e\bB,
\eeqa
together with the  duality-invariant
quantities $\bbJ_{e}  =  \frac{\displaystyle \bJ_{e}}{\displaystyle  
e},
\brho_{c}  =  \frac{\displaystyle \rho_{c}}{\displaystyle e} $ and
$\bsig_{e}  =  \frac{\displaystyle \sigma_{e}}{\displaystyle e^{2}},$
then in terms of these fields, the equations of dyonic MHD can be  
written as
(dropping the $ ' $ on  $P$ and  $\rho_M $):
\beqa\label{eq:4.20}
\pa_{t}\rho_{M}+\grad\cdot\bv & = & 0, \qquad
\pa_{t}\brho_{c}+\grad\cdot\bbJ_{e}  = 0, \cr
\rho_{M}(\frac{\pa\bv}{\pa t}+\bv\cdot\grad \bv) & = &
-\grad P + \brho_{c}\cE
-\frac{1}{c}\bbJ_{e}\times\cB, \cr
\frac{m_{d}m_{di}}{\rho_{M}}\frac{\pa\bbJ_{e}}{\pa t} & = &
\frac{m_{d}}{2\rho_{M}}\grad P
+ \cE - \frac{1}{c}\bv\times\cB\mbox{} +
\frac{m_{d}}{\rho_{M}c}\bbJ_{e}\times\cB -
\frac{1}{\bsig_{e}}\bbJ_{e}, \cr
\grad\cdot\cB & = & 0, \qquad
\grad\cdot\cE = \brho_{c}(e^{2}+g^{2}), \cr
\grad\times\cE & = & \frac{1}{c}\frac{\pa\cB}{\pa t}, \qquad
\grad\times\cB  =  -\frac{4\pi}{c}(e^{2}+g^{2})\bbJ_{e} -
\frac{1}{c}\frac{\pa\cE}{\pa t}.
\eeqa
We note that, up to factors of $(e^{2}+g^{2})$, these are
exactly the
equations of ordinary MHD with suitable identifications.

In view of the exact analogy with ordinary MHD, we follow
the standard
analysis in going over to the ideal limit. The important
additional
ingredient here is that all approximations must preserve duality, and  
as we
shall discuss below there is more than
one way of doing so.
In the limit of low temperature and low frequency Ohm's law
reduces to
\beq\label{eq:4.21}
\bbJ_{e}=\bsig_{e}(\cE-\frac{1}{c}\bv\times\cB),
\eeq
so that as $\bsig_{e}\rta\infty$, the requirement that
$\bbJ_{e}$ remains
finite yields
\beq\label{eq:4.22}
\cE=\frac{1}{c}\bv\times\cB.
\eeq
This relation can be used to eliminate $\cE$ from the
equations. Now we have to consider how to implement  the
low frequency limit in the last two equations
of the dualized Maxwell equations (\ref{eq:3.1}).
{}From these equations it might seem reasonable to drop both
$\frac{\displaystyle \pa\bE}{\displaystyle \pa t}$ and
$\frac{\displaystyle \pa\bB}{\displaystyle \pa t}$ in comparison
to the currents $\bJ_e $ and $\bJ_m $ respectively. However
whilst this certainly preserves duality, it leads to a constrained
effective ideal MHD in which $\bE $ and $\bB $ are forced to vanish.
This is a consequence of the fact that these fields arise purely
from the currents $\bJ_e $ and $\bJ_m $, and since the latter are
proportional
then $\bE $ and $\bB $ are parallel.  This condition is incompatible
with (\ref{eq:4.22}) unless  $\bE = \bB = 0 $.

There is an alternative approach however. Clearly we can apply the
usual low frequency approximation to the MHD equations, $i.e.$,
we can  neglect the
$\frac{\displaystyle \pa\cE}{\displaystyle \pa t}$ term in
Maxwell's equation for $\grad\times\cB$, and keep
the term $\frac{\displaystyle \pa\cB}{\displaystyle \pa t}$ in the
equation for $\grad\times\cE$. This also manifestly
preserves duality. However unlike the
case of usual MHD, in terms of the physical fields $\bE $ and $\bB$
such an approximation does not mean that the $t$ variation
are negligible compared to the currents, indeed they are
comparable which is why they are not neglected in
(\ref{eq:3.1}). Nevertheless the {\it combined}  $t$ variation
of these fields in $\cE $ given in (\ref{eq:4.18})
is assumed negligible compared to the currents, so a kind
of cancellation must occur.  Finally, the
infinite conductivity
limit $\bsig_{e}\rta\infty$ means that $\brho_{c}\rta 0$.

Implementing these
approximations and after finally taking the curl of the
force equation to
remove the $\grad P$ term, we obtain:
\beqa\label{eq:4.23}
\grad\cdot\bv & = & 0, \qquad\grad\cdot\cB  =  0 \cr
\grad\times (\bv\times\cB) & = & \frac{\pa\cB}{\pa t}, \qquad
\grad\times(\frac{\pa\bv}{\pa t}+\bv\cdot\grad \bv)  =
\frac{1}{4\pi\rho_{M}(e^{2}+g^{2})}
\times (\cB\cdot\grad \cB)
\eeqa
Defining $\bar{\cB} \equiv \frac{ 1}{ \sqrt{e^{2}+g^{2}}} \cB$
the above ideal dyonic MHD equations are exactly those of
ordinary ideal
MHD
with the identifications $\bE\lra\cE$ and $\bB\lra\bar{\cB}$, or  
equivalently
to those of dual MHD with  $\bE\lra -\bar\cB$ and $\bB\lra {\cE}$.

This is an interesting and unexpected result, and it is perhaps
worthwhile to try
and understand it further, particularly when we compare this to the
situation in the previous section. It seems that by forcing the
electric and magnetic charges to be localized simultaneously at a
point ($i.e.$ the case of dyons), we avoid the
 ``over constrained" system of equations arising from
the Ohm's laws (\ref{eq:3.21}), which led to the vanishing of
$\bE$ and $\bB$. Moreover we can try
to understand the dyonic case by considering the conserved  
generalized
kinetic energy  ${ E}_k$ of the ideal plasma in
the light of duality symmetry. This is given by
\beq\label{eq:4.24a}
{E}_k \, = \, \int d^3 x \{ \frac{\rho_M}{2} \bv \cdot \bv +
 \frac{\cB \cdot \cB }{ 4 \pi  (e^2 + g^2 ) } \}.
\eeq
which is easily seen to be duality invariant. Moreover this  
expression
for ${E}_k $ goes over to the energy we would expect in a plasma
of purely electric or magnetic charges as we send either $g  
\rightarrow 0$
or  $e \rightarrow 0$ respectively. Now if we again believe that  
duality
symmetry in our system is independent of the values of the fluid  
velocity
$\bv $ then consider  going to a frame which is comoving
with respect to the fluid , $i.e.$, the limit $\bv \rightarrow 0 $.  
Now,
crucially it is the
duality invariant field $\cE $ that vanishes according to  
(\ref{eq:4.22})
not the physical electric field itself, which was the case for
the plasma of separate $e$ and $g$ charges considered earlier.
If either $\bE $ or $\bB$ vanishes, then unbroken duality symmetry
implies they both vanish.
In fact $\cE \, = \, 0 $ implies that
$\bE \, = \, - \frac{\displaystyle g}{\displaystyle e} \bB $, $i.e.$,  
the
electric
and magnetic fields are anti-parallel (in our conventions $g$ and $e$  
have the
same signs), so that both $\bE $ and $\bB $ are non-vanishing in this  
limit,
and consequently contribute to the energy $E_k $ in a duality
invariant way.

To summarize, we have considered the generalizations of the theory of
MHD to both the cases of a plasma of separate magnetic and electric  
charges and
a
 dyonic plasma. In the limit of infinite
electric and magnetic conductivities, it was found that
 only in the case of dyonic plasmas can one have  non-vanishing  
magnetic and
electric fields.
 By considering all fields to be independent of one of the
coordinates, one can obtain an effectively 2d  theory of  duality  
invariant MHD
which
also includes passive scalars. The application of CFT techniques to  
the study
of turbulence
 in these models, which extends the investigations in [2,3],
 is considered in a separate publication [4].
Finally, the ideas of [5] in which a string (magnetic and electric)  
was
proposed as a solution to the
 equations of MHD (and its dual) can be generalized to the duality  
invariant
case
corresponding to a  dyonic plasma.
This will be reported on in a forthcoming paper.

\vskip0.3in
\begin{center}
{\bf\Large Acknowledgements}
\end{center}
\vskip0.1in
The work of S.Thomas  was supported by the
Royal Society of Great Britain
and  W. A. Sabra  by PPARC.
\newpage
\begin{center}
\noin{\bf\Large References}
\end{center}
\vspace{2ex}
\begin{description}
\item{[1]} A .M. Polyakov {\it Nucl. Phys.} {\bf B396} (1993) 367.
\item{[2]} G. Ferretti and Z. Yang, {\it Europhys. Lett.} {\bf 22}   
(1993) 639.
\item{[3]} O. Coceal and S. Thomas,
{\it Conformal Models of Magnetohydrodynamic Turbulence},
preprint QMW-PH-95-45, hep-th/9512022.
\item{[4]} O. Coceal, W. A. Sabra and S.Thomas,
{\it Conformal solutions of duality invariant 2d Magnetohydrodynamic
Turbulence}, QMW-PH-96-05.
\item{[5]} P. Olesen, {\it Dual strings and Magnetohydrodynamics},
NBI-HE-95-31,hep-th9509023.
\item{[6]} D. R. Nicholson, {\it Introduction to Plasma Theory},  
Wiley (New
York) 1983.
\item{[7]} P. Goddard and D. I. Olive, {\it Magnetic monopoles in  
gauge field
theories}, Rep. Prog. Phys. {\bf 41} (1978) 1357.
\item{[8]} B. Nienhuis, {\it J. Stat. Phys.} {\bf 34} (1984) 731.
\item{[9]} Y. Aharonov and D. Bohm, {\it Phys. Rev.}{\bf 115}, 484  
(1959).
\item{[10]} J. L. Cardy and E. Rabinovici, {\it Nucl. Phys.} {\bf  
B205} (1981)
1.
\end{description}
\end{document}